\documentclass[]{agujournal2018}
\usepackage{url} %this package should fix any errors with URLs in refs.
\draftfalse

\newcommand{\adv}{    {\it Adv. Space Res. }}

\newcommand{\aap}{    {\it Astron. Astrophys. }}

\newcommand{\apj}{    {\it Astrophys. J. }}
\newcommand{\apjl}{   {\it Astrophys. J. Lett. }}
\newcommand{\apjs}{   {\it Astrophys. J. Suppl. S.}}

\newcommand{\grl}{    {\it Geophys. Res. Lett. }}

\newcommand{\jgr}{    {\it J. Geophys. Res. }}

\newcommand{\nat}{    {\it Nature }}

\newcommand{\solphys}{{\it Solar Phys. }}

\newcommand{\ssr}{    {\it Space Sci. Rev. }}

\newcommand{\planss}{   {\it Planet Space Sci. }}

\journalname{Space Weather}

\begin{document}

%% ------------------------------------------------------------------------ %%
%  Title
%
% (A title should be specific, informative, and brief. Use
% abbreviations only if they are defined in the abstract. Titles that
% start with general keywords then specific terms are optimized in
% searches)
%
%% ------------------------------------------------------------------------ %%

% Example: \title{This is a test title}

\title{Evaluation of CME Arrival Prediction Using Ensemble Modeling Based on Heliospheric Imaging Observations}

%% ------------------------------------------------------------------------ %%
%
%  AUTHORS AND AFFILIATIONS
%
%% ------------------------------------------------------------------------ %%

% Authors are individuals who have significantly contributed to the
% research and preparation of the article. Group authors are allowed, if
% each author in the group is separately identified in an appendix.)

% List authors by first name or initial followed by last name and
% separated by commas. Use \affil{} to number affiliations, and
% \thanks{} for author notes.
% Additional author notes should be indicated with \thanks{} (for
% example, for current addresses).

% Example: \authors{A. B. Author\affil{1}\thanks{Current address, Antartica}, B. C. Author\affil{2,3}, and D. E.
% Author\affil{3,4}\thanks{Also funded by Monsanto.}}

\authors{Tanja Amerstorfer\affil{1}, J\"{u}rgen Hinterreiter\affil{1,2}, Martin A. Reiss\affil{1,3}, Christian M\"{o}stl\affil{1,3}, Jackie A. Davies\affil{4}, Rachel L. Bailey\affil{1,5}, Andreas J. Weiss\affil{1,2,3}, Mateja Dumbovi\'{c}\affil{6}, Maike Bauer\affil{1,2}, Ute V. Amerstorfer\affil{1}, Richard A. Harrison\affil{4}}

% \affiliation{1}{First Affiliation}
% \affiliation{2}{Second Affiliation}
% \affiliation{3}{Third Affiliation}
% \affiliation{4}{Fourth Affiliation}

\affiliation{1}{Space Research Institute, Austrian Academy of Sciences, Schmiedlstra\ss e 26, 8042 Graz, Austria}
\affiliation{2}{Institute of Physics, University of Graz, Universit\"{a}tsplatz 5/II, 8010 Graz, Austria}
\affiliation{3}{Institute of Geodesy, Graz University of Technology, Steyrergasse 30, 8010 Graz, Austria}
\affiliation{4}{RAL Space, Rutherford Appleton Laboratory, Harwell Campus, Didcot OX11 0QX, UK}
\affiliation{5}{Conrad Observatory, Zentralanstalt f\"{u}r Meteorologie und Geodynamik, Vienna, Austria}
\affiliation{6}{Hvar Observatory, Faculty of Geodesy, University of Zagreb, Zagreb, Croatia}
%(repeat as many times as is necessary)

%% Corresponding Author:
% Corresponding author mailing address and e-mail address:

% (include name and email addresses of the corresponding author.  More
% than one corresponding author is allowed in this LaTeX file and for
% publication; but only one corresponding author is allowed in our
% editorial system.)

% Example: \correspondingauthor{First and Last Name}{email@address.edu}

\correspondingauthor{Tanja Amerstorfer}{tanja.amerstorfer@oeaw.ac.at}

%% Keypoints, final entry on title page.

%  List up to three key points (at least one is required)
%  Key Points summarize the main points and conclusions of the article
%  Each must be 100 characters or less with no special characters or punctuation

% Example:
% \begin{keypoints}
% \item List up to three key points (at least one is required)
% \item Key Points summarize the main points and conclusions of the article
% \item Each must be 100 characters or less with no special characters or punctuation
% \end{keypoints}

\begin{keypoints}
\item CME prediction tool ELEvoHI is ready to be used in real time, based on STEREO-A/HI beacon data
\item Different model set-ups and inputs lead to large differences of the prediction accuracies
\item Accurate modeling of the ambient solar wind is of particular importance to improve CME predictions
\end{keypoints}

%% ------------------------------------------------------------------------ %%
%
%  ABSTRACT
%
% A good abstract will begin with a short description of the problem
% being addressed, briefly describe the new data or analyses, then
% briefly states the main conclusion(s) and how they are supported and
% uncertainties.
%% ------------------------------------------------------------------------ %%

%% \begin{abstract} starts the second page

\begin{abstract}
In this study, we evaluate a coronal mass ejection (CME) arrival prediction tool that utilizes the wide-angle observations made by STEREO's heliospheric imagers (HI). The unsurpassable advantage of these imagers is the possibility to observe the evolution and propagation of a CME from close to the Sun out to 1 AU and beyond. We believe that by exploiting this capability, instead of relying on coronagraph observations only, it is possible to improve today's CME arrival time predictions.
The ELlipse Evolution model based on HI observations (ELEvoHI) assumes that the CME frontal shape within the ecliptic plane is an ellipse, and allows the CME to adjust to the ambient solar wind speed, i.e.\ it is drag-based. ELEvoHI is used to perform ensemble simulations by varying the CME frontal shape within given boundary conditions that are consistent with the observations made by HI. In this work, we evaluate different set-ups of the model by performing hindcasts for 15 well-defined isolated CMEs that occurred when STEREO was near L4/5, between the end of 2008 and the beginning of 2011. In this way, we find a mean absolute error of between $6.2\pm7.9$ h and $9.9\pm13$ h depending on the model set-up used.
ELEvoHI is specified for using data from future space weather missions carrying HIs located at L5 or L1. It can also be used with near real-time STEREO-A HI beacon data to provide CME arrival predictions during the next $\sim7$ years when STEREO-A is observing the Sun-Earth space.
\end{abstract}

%% ------------------------------------------------------------------------ %%
%
%  TEXT
%
%% ------------------------------------------------------------------------ %%

%%% Suggested section heads:
% \section{Introduction}
%
% The main text should start with an introduction. Except for short
% manuscripts (such as comments and replies), the text should be divided
% into sections, each with its own heading.

% Headings should be sentence fragments and do not begin with a
% lowercase letter or number. Examples of good headings are:

% \section{Materials and Methods}
% Here is text on Materials and Methods.
%
% \subsection{A descriptive heading about methods}
% More about Methods.
%
% \section{Data} (Or section title might be a descriptive heading about data)
%
% \section{Results} (Or section title might be a descriptive heading about the
% results)
%
% \section{Conclusions}

\section{Introduction} \label{sec:intro}

As the main drivers of space weather events, coronal mass ejections (CMEs) are one of the most important subjects to be investigated as part of current solar-terrestrial research. CMEs are impulsive outbursts of the solar corona, consisting of a magnetic flux rope that impounds coronal material and solar wind particles during its propagation through the interplanetary medium. Fast CMEs can reach speeds of up to $3000$ km~s$^{-1}$ and, depending on their speeds and the characteristics of their intrinsic magnetic fields, can cause, for example, severe issues for satellites and disruptive geomagnetic disturbances at Earth \citep{wil87,tsu88,hut05,far06,gop08}. One of the most difficult CME properties to predict is the orientation of the magnetic field inside the CME, which is, at the same time, the most critical parameter due to the fact that a large southward magnetic field component facilitates the strongest geomagnetic storms. A large number of studies are currently tackling this task by developing new models that try to predict the orientation of the magnetic field at 1~AU \citep[e.g.][]{sav15,kub16,shi16,kaygop17,pal17,moe18,ver19,sin20}.

Besides the magnetic field, the arrival speed of the CME plays an important role as high impact speeds, including those of the shock-front driven by the CME, can intensify a geomagnetic disturbance \citep{gos91,yue10,oli18}. Generally, geoeffectiveness is related to the dawn-to-dusk electric field and therefore to the flow speed \citep[][]{obrmcp00}. While prediction of the orientation of the magnetic field within a CME is particularly difficult---especially due to the lack of magnetic field measurements in the corona---the prediction of the CME arrival time and speed can be carried out using different kinds of data and numerous prediction models. In particular, accurate prediction of the shock arrival time at Earth is crucial in order to be able to react accordingly to an expected disturbance. However, the timing and the probability of arrival at Earth are both still hard to predict. \citet{wol18} analyzed the real-time predictions performed at the Community Coordinated Modeling Center (CCMC) using the WSA-ENLIL+Cone model between the years 2010 and 2016. They found that the success ratio, reflecting the fraction of correct predictions, to be 0.4 and the false alarm ratio to be 0.6. This demonstrates the necessity of improving arrival time and probability prediction of CMEs.

Most prediction models rely on images from coronagraphs that observe the solar corona out to a maximum plane-of-sky distance of 30 R$_\odot$ \citep[e.g.][]{dum18,sin18,plu19,kay20}. The big advantages of these observations are their availability in real-time and their relatively simple interpretation. In coronagraph images, the inferred distances can be directly used without any consideration of Thomson scattering effects, which is commonly known as the plane-of-sky assumption. Additionally, the integration of the scattered photospheric light along the line-of-sight can be neglected, since the extent of a CME is rather small close to the Sun. The big drawback is the small field of view that corresponds to a maximum one seventh of the Sun-Earth distance. \cite{ril18} analyzed the accuracy of models contributing to the CME scoreboard\footnote{\url{https://kauai.ccmc.gsfc.nasa.gov/CMEscoreboard}}, a platform that is used by scientists and model developers to test their models in real-time. It was found that the model with the best performance (WSA-ENLIL+Cone run at NOAA/SPWC) achieved a mean absolute arrival time error of 13 h with a standard deviation of $\pm 15$ h. The predictions evaluated were made in real-time over a time range of almost 6 years, i.e.\ the numbers in that study reflect the state of the art better than any of the other studies that covered only a small number of selected events.

Other instruments that enable CMEs to be observed in white-light are the heliospheric imagers \citep[HI;][]{eyl09} on-board the Solar TErrestrial RElations Observatory \citep[STEREO;][]{kai08}. These wide-angle cameras image the space between the Sun and 1 AU and beyond. Due to their large field of view, line-of-sight integration is an important factor when interpreting these images and the plane-of-sky assumption is not valid for HI. Therefore, it is necessary to assume a certain longitudinal extent of the CME frontal shape, as well as being aware that it is not possible to follow the same part of the CME front during its propagation through the entire field of view of HI. One of the drawbacks of STEREO/HI data is that the near real-time beacon data suffer from low time and spatial resolution and from data gaps, i.e.\ it is expected that real-time predictions based on HI beacon data cannot achieve the same accuracy as predictions based on HI science data \citep[][]{tuc15}. Now that STEREO-A is again observing the space between Sun and Earth from an optimal vantage point, predictions using HI beacon data will no doubt be contributed to the CME scoreboard in the future. ESA is currently planning a space weather mission to the observationally advantageous Lagrange point 5 (L5) of the Sun-Earth system, located around 60$^\circ$ behind the Sun-Earth line \citep[][]{gib17}. This mission is dedicated to space weather prediction and will, if funded, carry HI cameras providing real-time data with quality comparable to STEREO/HI science data. This could be an important step forward to improving CME arrival time and speed prediction.

With regard to this and other possible future space weather monitoring missions carrying heliospheric imagers, we present a detailed evaluation of different model parameters and inputs to the ELlipse Evolution model based on single spacecraft HI observations \citep[ELEvoHI;][]{rol16,ame18}. ELEvoHI is designed to be operational in real-time as soon as HI real-time data are available with sufficient quality to be used by this model. We have found that small changes within the model, its parameters or inputs, can lead to a large difference in the CME arrival prediction. In the following sections, we investigate different ways of using ELEvoHI together with HI science data and compare these approaches to each other in order to identify the optimal model set-up leading to the smallest prediction errors in time and speed.

\section{Data} \label{sec:data}

We use a list of 15 well-observed (remotely and in situ) non-interacting Earth-directed CMEs within the time range extending from the end of 2008 until the beginning of 2011 (Table \ref{tab:table0}). During this time, STEREO was in an ideal location (between 45 and 90$^\circ$ east and west of Earth) to observe Earth-directed events. Unfortunately, due to low solar activity during these years, the number of fast CMEs in this interval is very small, i.e.\ only one event arrived at Earth with a speed of more than 700 km~s$^{-1}$, while most of the events in the list were detected in situ with a speed of less than 400 km~s$^{-1}$.

Parts of this study use coronagraph images provided by (1) the SOHO mission, with LASCO C2 and C3 \citep[][]{bru95}, which observe the space around the Sun between 2 and 30 R$_\odot$ in the plane of sky, and by (2) STEREO from two different vantage points, with COR2 \citep[][]{how08} having a field of view extending from 2 to 15 R$_\odot$. For parts of this study, we use coronagraph observations from all three vantage points together to get an estimate of the CME shape. The most important data source for this study and the ELEvoHI model are provided by the heliospheric imagers on-board STEREO. The HI instrument on each spacecraft consists of two white-light wide-angle cameras: HI1 having an angular field of view in the ecliptic of 4--24$^\circ$ from Sun-center and HI2 having an angular field of view, again in the ecliptic, of 18--88$^\circ$, roughly corresponding to a heliocentric distance of 1~AU. For this study, we used HI science data, having a time-cadence of 40 minutes (HI1) and 2 hours (HI2). Three events in the list (n$^\circ$ 4\&5, 9\&10, 14\&15) are observed from STEREO-A and STEREO-B. CMEs viewed from the two different spacecraft are treated separately, i.e.\ are not combined into a single prediction.

In order to evaluate the prediction accuracy of ELEvoHI, arrival times and speeds given in the ICMECAT \citep[][]{moe17} catalog provided by the ''Heliospheric Cataloguing, Analysis and Techniques Service`` (HELCATS) project\footnote{\url{www.helcats-fp7.eu}} were used. This catalog lists, among those for other spacecraft, the interplanetary CME (ICME) shock arrivals detected by the Wind spacecraft \citep[][]{lep95,ogi95} located at L1. Parts of this study rely on information about the solar wind speed at 1~AU detected by the Wind spacecraft, that is used as approximation for the ambient solar wind speed influencing the CME during its propagation (Section \ref{sec:solarwind}).

\begin{table}[h!]
\caption{Overview of events used in this study. The columns state the event number, the time and date when the CME was first visible in HI1, the observing STEREO spacecraft, the HEE longitude (i.e.\ the separation of the observing spacecraft from Earth), the in situ arrival time and speed detected at Earth. This information was taken from the HELCATS project website.}
\label{tab:table0}
\centering
\begin{tabular}{cccccc}
\hline
n$^\circ$ & First observed by HI1 [UT] & HI observer & HEE longitude [$^\circ$] & Arrival time [UT] & Arrival speed [km~s$^{-1}$]\\
\hline
1 & 2008-12-12 15:29 & A & 42.3 & 2008-12-17 03:35 & 355 \\
2 & 2009-01-30 20:09 & B & -46.1 & 2009-02-03 19:11 & 360 \\
3 & 2009-09-03 23:29 & A & 59.6 & 2009-09-10 10:19 & 306 \\
4 & 2010-02-03 14:49 & A & 64.7 & 2010-02-07 18:04 & 406 \\
5 & 2010-02-03 20:49 & B & -70.7 & 2010-02-07 18:04 & 406 \\
6 & 2010-03-19 20:09 & B & -71.4 & 2010-03-23 22:33 & 292 \\
7 & 2010-04-03 12:09 & A & 67.5 & 2010-04-05 07:55 & 734 \\
8 & 2010-04-08 06:49 & A & 67.8 & 2010-04-11 12:28 & 432 \\
9 & 2010-05-23 22:09 & A & 71.6 & 2010-05-28 02:23 & 370 \\
10 & 2010-05-24 00:09 & B & -70.0 & 2010-05-28 02:23 & 370 \\
11 & 2010-06-16 23:29 & B & -69.8 & 2010-06-21 03:35 & 401 \\
12 & 2010-10-26 16:10 & B & -80.6 & 2010-10-31 02:09 & 366 \\
13 & 2010-12-15 21:29 & A & 85.2 & 2010-12-19 20:23 & 381 \\
14 & 2011-01-30 20:09 & A & 86.2 & 2011-02-04 01:55 & 376 \\
15 & 2011-01-30 18:49 & B & -93.0 & 2011-02-04 01:55 & 376 \\
\hline
%\multicolumn{2}{l}{$^{a}$Footnote text here.}
\end{tabular}
\end{table}

\section{ELEvoHI at a glance} \label{sec:elevohi}

%\subsection{ELlipse Evolution based on HI} \label{sec:elevohi}

The ELlipse Evolution model based on Heliospheric Imager data (ELEvoHI) was first presented by \cite{rol16} as a single-run model, where it was shown that including solar wind drag leads to an improvement of CME arrival time and speed predictions over the common HI prediction methods, such as Fixed-Phi \citep[][]{rou08,kahweb07}, Harmonic Mean \citep[][]{howtap09,lug09b} or Self-similar Expansion fitting \citep[][]{moedav13,dav12}. Allowing the CME to adjust its kinematics to the ambient solar wind flow particularly improves the arrival speed predictions, which has direct relevance to accurately predicting geomagnetic storm strength \citep[][]{rol16}.

\citet{ame18} introduced the ELEvoHI ensemble approach and tested it using a case study, in which a CME was detected in situ by two radially aligned spacecraft at 0.48 and 1.08 AU. The authors showed that it is possible to predict CME arrival at the observing spacecraft itself, i.e.\ it is possible to predict a halo CME, supporting the idea of having an HI instrument positioned at L1.

\begin{figure}[h]
    \centering
    \includegraphics[]{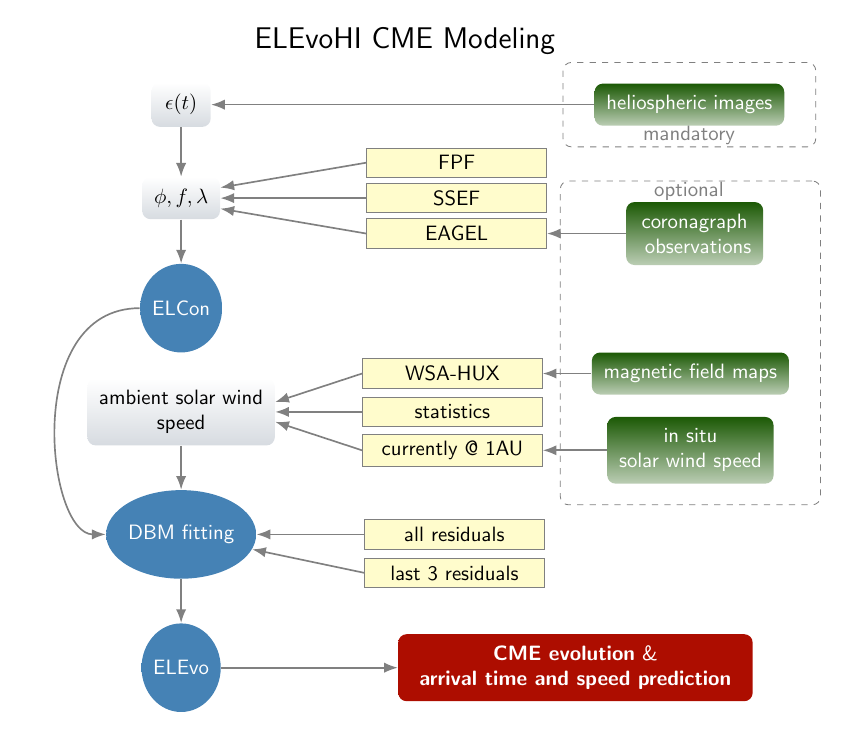}
    \caption{Schematic illustration of all parts contributing to an ELEvoHI ensemble prediction. The green boxes are the possible data used, the blue boxes are the three main modules building ELEvoHI. The gray boxes are all input parameters needed. The yellow boxes show the possible sources of these input parameters grouped into three different parts that are tested in this study. The red box is the model output, i.e.\ kinematical profiles and arrival time and speed predictions at the target of interest.}
    \label{fig:flowchart}
\end{figure}

ELEvoHI is a combination of three main modules that derive parameters from observations to serve as input to the next module. Figure \ref{fig:flowchart} presents the prediction scheme based on ELEvoHI ensemble modeling used in this paper. The left column shows different inputs (gray boxes) to the three main modules of ELEvoHI (blue ellipses), resulting in the modeling and prediction results (red box). The green boxes on the right show the different data that can be used to drive the model, while only data from heliospheric imagers is mandatory and all other data are optional. The middle part of the figure (yellow boxes) presents the three groups of inputs that this study investigates in order to identify their best combination (in terms of CME geometry, ambient solar wind speed, and DBM fitting). In the following paragraphs, the individual steps within ELEvoHI (blue circles in Figure \ref{fig:flowchart}) in its ensemble approach are briefly described:

The starting point is the CME time-elongation track, $\epsilon(t)$, acquired from HI observations, usually from a time-elongation map at fixed position angle. This track is converted from angular units to units of radial distance by ELEvoHI's built-in procedure ELlipse Conversion (ELCon), based on an ensemble of assumed front shapes and propagation directions (see below). Detailed information on the ELCon conversion method can be found in \citet{rol16}.

In the next step, each ensemble member time-distance track for the CME is fitted using a equation of motion based on the drag-based model (DBM) given in \cite{vrs13}:

\begin{equation}
R(t) = \pm 1/\gamma \ln[1 \pm \gamma(v_{\rm{init}} - w)t] + wt + r_{\rm{init}},
\end{equation}
where $r_{\rm{init}}$ is the initial distance and $v_{\rm{init}}$ the initial speed.
The sign $\pm$ is positive when the CME is accelerating ($v_{\rm{init}} < w$) and negative when it is decelerating ($v_{\rm{init}} > w$) due to the drag-force exerted by the ambient solar wind. The drag parameter, $\gamma = C_{\rm{D}} \frac{A_{\rm{CME}} \rho_{\rm{sw}}}{m_{\rm{CME}}}$, is the parameter that results from least-square fitting of the time-distance track within the DBM fitting routine implemented in ELEvoHI. $C_{\rm{D}}$ is the drag-coefficient assumed to equal 1, $A_{\rm{CME}}$ is the CME cross section that the drag is acting on, $m_{\rm{CME}}$ is the CME mass, and $\rho_{\rm{sw}}$ is the solar wind density.
Within the DBM fitting procedure, $t_{\rm{init}}$, the initial time of the fit, is defined manually by the user once for each event. Subsequently, $r_{\rm{init}}$ and $v_{\rm{init}}$ are derived separately for each ensemble member from the output of ELCon.

\begin{figure}[h]
    \centering
    \includegraphics[width=15cm,keepaspectratio]{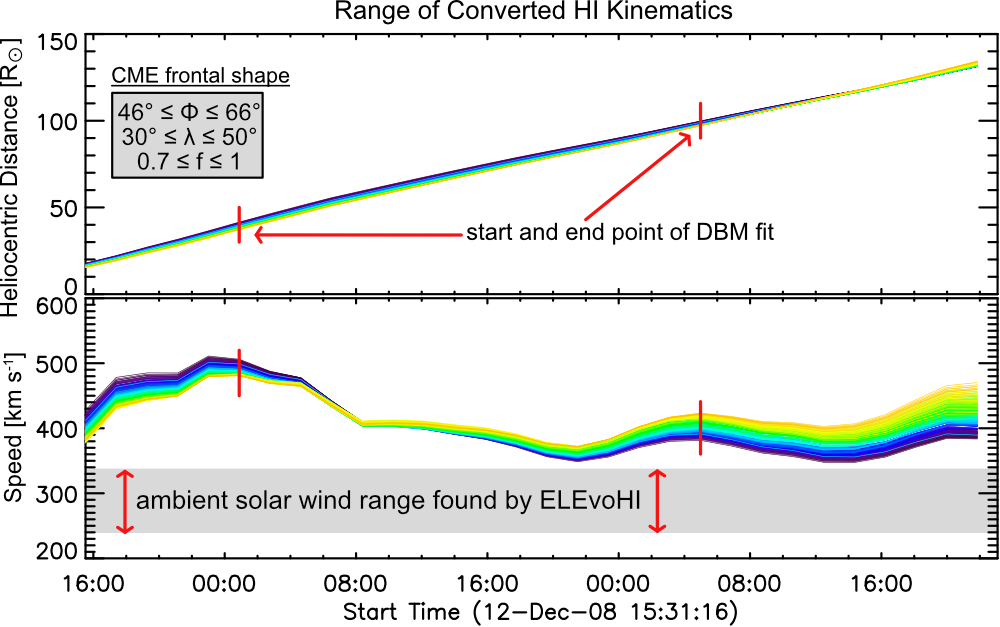}
    \caption{Range of HI kinematics (upper panel: time-distance profiles, lower panel: time-speed profiles) resulting from the input parameters corresponding to different CME frontal shapes and directions. The red vertical lines mark the start and end times of the HI data used for CME arrival prediction, the gray shaded area in the lower panel illustrates the range of the ambient solar wind speed resulting from drag-based fitting to the HI kinematics.}
    \label{fig:elcon}
\end{figure}

The procedure of defining the ambient solar wind speed, $w$, is described in Section \ref{sec:solarwind}. Figure \ref{fig:elcon} demonstrates the approach of ELCon and the following DBM fitting for one example CME (CME n$^\circ$ 1 in Table \ref{tab:table0}). The upper panel shows the time-distance profile derived from the STEREO-A HI time-elongation track by using 220 different combinations of frontal shape-related input parameters (angular half width, $\lambda$, and inverse ellipse aspect ratio, $f$) and propagation direction, $\phi$. Each of these three parameters is varied within a certain range to build an ensemble of different CME shapes, from each of which a prediction is made. Depending on the assumed angular width, aspect ratio and direction of the tracked feature, the derived kinematics differ for each ensemble member. The lower panel shows the interplanetary speed profiles of the CME apex derived by ELCon from each of the time-distance profiles. The red vertical lines mark the start and the end point of the HI data used for DBM fitting (fits are not shown) and for making the CME arrival prediction.

The parameters obtained by DBM fitting serve as input for the ELlipse Evolution model \citep[ELEvo;][]{moe15} that produces the arrival prediction. ELEvo runs the DBM by propagating the previously-defined elliptical CME frontal shape in the also previously defined direction, which is different for each ensemble member, and predicts its arrival at any target of interest based on the drag parameter and ambient solar wind speed derived from DBM fitting.

In the following, we describe the different methods used to derive input parameters for ELEvoHI, such as information on the CME frontal shape, propagation direction and the ambient solar wind speed.
All of them are optional and can be replaced by a basic statistical estimation or a simple assumption.

\section{Deriving the input parameters for ELEvoHI}

\subsection{Direction, angular width and curvature of the CME front}
%\subsection{Graduated Cylindrical Shell fitting} \label{sec:gcs}

Besides the time-elongation track measured from HI observations, ELEvoHI needs information on the frontal shape, i.e.\ $f$ and $\lambda$, and the direction of motion of the CME. The latter can either be gained from HI observations or from coronagraph observations, which additionally provide the possibility to estimate the angular width.

\subsubsection{Ecliptic cut Angles from GCS for ELEvoHI}\label{sec:eagel}

The first potential method to provide $\phi$ and $\lambda$ parameters used by ELEvoHI is based on the Graduated Cylindrical Shell fitting method \citep[GCS fitting;][]{the06,the09,the11}. GCS fitting (implemented within SolarSoft, \texttt{rtsccguicloud.pro}) enables the manual fitting of a croissant-shaped CME body to simultaneous images from coronagraphs observing from different vantage points. In our study, we use images from STEREO/COR2 from both sides, as well as LASCO/C2 and/or C3 images. Several shape-related CME parameters can be adjusted within a widget tool until the best match with the CME visible within the coronagraph images is achieved. For our purposes, GCS is run as a part of the so-called EAGEL (Ecliptic cut Angles from GCS for ELEvoHI) tool, which is described below.

\begin{figure}[h]
    \centering
    \includegraphics[width=15cm,keepaspectratio]{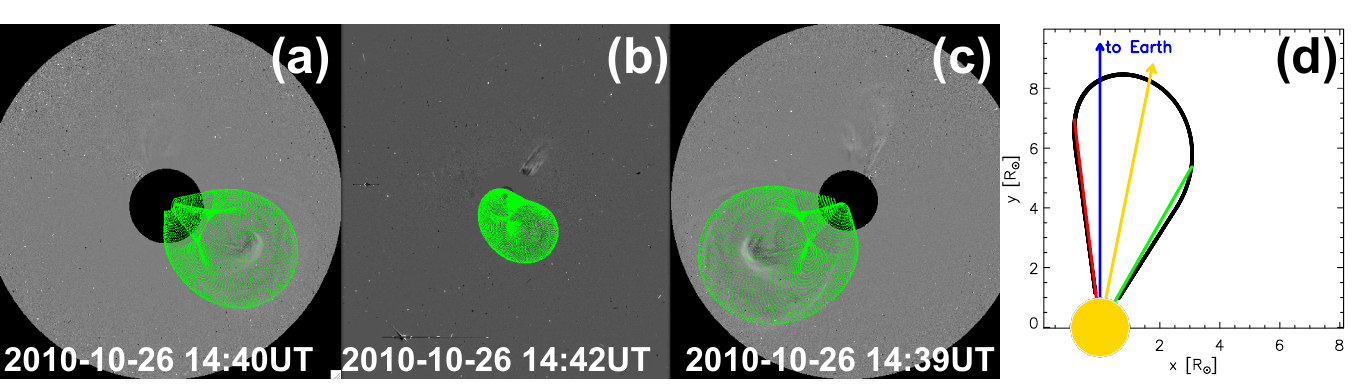}
    \caption{Example of a CME (event n$\circ$ 12) having a large inclination relative to the ecliptic plane. Panels a)---c) show the GCS reconstruction of the CME shape overlaid on STEREO-A/COR2, SOHO/C3 and STEREO-B/COR2 difference images. Panel d) shows the ecliptic cut resulting from the EAGEL tool, with the blue arrow pointing towards Earth, the green and red lines defining the outer edges of the CME and the yellow arrow pointing along the direction of motion used as input to ELEvoHI.}
    \label{fig:eagel}
\end{figure}

Within EAGEL the download and pre-processing of the coronagraph data is included in such a way that a CME is clearly recognizable in the images. Based on these images, GCS fitting of a CME is performed. EAGEL then creates an ecliptic cut of the wire-frame of the fitted CME and calculates $\lambda$ and $\phi$ with respect to Earth, STEREO-A and STEREO-B. ELEvoHI is operated in an ensemble mode, in which the input values of shape and direction are varied within a pre-defined range. In the case that inputs from EAGEL are used, $\lambda$ and $\phi$ are each varied within $\pm 10^\circ$. This range is chosen based on a previous study by \citet{mie10}, who cite this as the error range of these parameters when different observers manually fit the same CME using GCS.
Panels a)--c) in Figure \ref{fig:eagel} show a GCS fit to one of the CMEs under study (n$^\circ$ 12 in Table \ref{tab:table0}). In case of this CME and due to the high tilt angle ($\approx -28^\circ$), the ecliptic cut conducted by the EAGEL tool corresponds to a very narrow structure as shown in panel d). Because of the $\pm 10^\circ$ in $\lambda$ and $\phi$ used in the ELEvoHI ensemble mode, the whole ensemble appears to be relatively wide compared to the input ecliptic cut. To build the ensemble, these inputs are varied using a step size of $\Delta \phi = 2^\circ$ and $\Delta \lambda = 5^\circ$. The parameter $f$, which is related to the curvature of the front, is not obtained from the ecliptic cut but is, instead, varied between 0.8 (flat elliptical frontal shape) and 1 (circular frontal shape).

\subsubsection{FPF and SSEF methods}

In the study predicting CME arrival times and speeds using ELEvoHI performed by \cite{rol16}, the propagation direction was obtained from FPF and the same angular half-width, namely $35^\circ$, was used for every CME in the list. Although this is a quick and easy approach with no additional need for coronagraph data, it does not provide information about the true angular half-width of the CME. With such information as input, we might be able to improve ELEvoHI's prediction accuracy. In the study by \cite{ame18}, the information on the CME frontal shape was taken from an intersection of the GCS shape with the ecliptic plane (as discussed in Section \ref{sec:eagel}). That case study resulted in a prediction with very high precision.

To test the effect of assuming a finite CME width, we use the direction of motion from Fixed-Phi Fitting \citep[FPF;][]{kahweb07,rou08} as well as from Self-Similar Expansion Fitting \citep[SSEF;][]{dav12,moedav13}. These methods are analogous except that, in the latter, the CME is not assumed to be a point and one has to assume an angular half-width for the circular shaped CME front. FPF and SSEF both perform a numerical fit to the time-elongation profile of the CME track measured from HI observations, hence they are based on the same input as ELEvoHI. Both methods assume a constant propagation direction and, in contrast to ELEvoHI, a constant propagation speed. We derive the propagation direction using SSEF assuming a half-width of $45^\circ$. The propagation directions from both HI fitting methods were then used together with a range of 30 -- 50$^\circ$ (and a step size of $5^\circ$) for the angular half-width within ELEvoHI.

\begin{figure}[h]
    \centering
    \includegraphics[width=15cm,keepaspectratio]{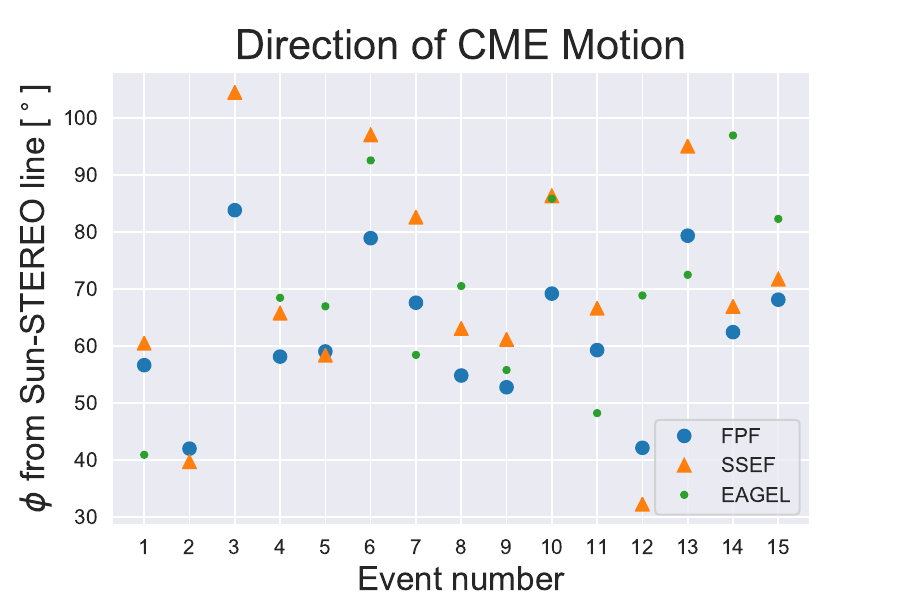}
    \caption{Absolute propagation directions relative to the Sun-STEREO line derived from EAGEL (green), FPF (blue) and SSEF (orange). For two events GCS fitting was not possible. The mean absolute difference of the resulting directions is around 12.7$^\circ$.}
    \label{fig:angle_comp}
\end{figure}

As a check, we compared the propagation directions resulting from FPF, SSEF and EAGEL for the 15 CMEs under study, and found that the mean absolute difference between the directions derived from the EAGEL approach and those from the HI fitting methods was around 14$^\circ$ and, between the two HI fitting methods, it was around 9$^\circ$. Figure \ref{fig:angle_comp} shows the derived directions of motion derived using the three methods (EAGEL: green dot, FPF: blue circle, SSEF: orange triangle) for each event studied. For events 2 and 3, no GCS fit could be performed due to the faint nature of the CME structure within the coronagraph images. Therefore, for these events, we have no prediction based on model set-ups using information from the EAGEL method. It is expected that the direction of motion and the angular half-width contribute significantly to the prediction accuracy. \citet{ame18} performed a sensitivity study that showed that, indeed, for the halo CME under study, the direction of motion had the biggest influence on the predicted transit time. However, this could be different for a side-on view of a CME or for different events.
It is important to emphasize that $\lambda$ and $\phi$ are the only parameters in our model that dictate if Earth (or any other target) is hit by the CME or not.

For the ELEvoHI model set-up test, as discussed in this section, we use the following inputs for the CME direction and angular half width:
\begin{enumerate}
\item EAGEL direction and half-width,
\item FPF direction and predefined angular half-width from $30-50^\circ$,
\item SSEF direction and predefined angular half-width from $30-50^\circ$.
\end{enumerate}

\subsection{Ambient solar wind speed}\label{sec:solarwind}

\subsubsection{In situ solar wind speed at 1 AU}

In its current version, ELEvoHI accepts only a constant (in space and time) background solar wind input. \cite{rol16} and \cite{ame18} assumed that the ambient solar wind at 1~AU is the same that influences the CME throughout its evolution, i.e.\ the solar wind speed at 1~AU was used as input to ELEvoHI. Note that a background solar wind speed prescribed in this way is not truly representative of the actual background wind through which the CME propagates. In that approach, the minimum and maximum solar wind speed values over the time range of the HI data (either from STEREO-A or B) are used for making the prediction, and three values in between those minimum and maximum values, as the basis for the DBM fitting. Hence, five DBM fits are performed, and the optimal fit (defined below) gives us the background speed, which is further used to perform the prediction.

\subsubsection{Statistical approach}

In order to find a better method, we investigate whether the DBM fit is able to `decide' for itself which solar wind speed best fits the CME kinematics. To this end, we calculated the mean solar wind speed in OMNI data between the years 2004 and 2018 to be $425$ km~s$^{-1}$ with a standard deviation of $100$ km~s$^{-1}$. We use these values to define the speed range utilized for DBM fitting as the mean value $\pm$ twice the standard deviation. For each ensemble member, we perform 17 DBM fits corresponding to speeds from 225 to 625 km~s$^{-1}$ in steps of 25 km~s$^{-1}$; the optimal DBM fit then yields the background solar wind speed.

This approach allows the model to select from a wide range of possible background solar wind speeds for itself. This is possible because the HI kinematics are not compatible with every possible solar wind speed. Depending on the CME speed and its evolution, i.e.\ whether the CME is decelerating, accelerating or propagating with a constant speed, only some candidate solar wind speeds will result in a converging DBM fit. Due to the wide range of ensemble members, each having different kinematics (see gray area in the lower panel of Figure \ref{fig:elcon}, the selected solar wind speed can be different for each ensemble member.

\subsubsection{Input from WSA-HUX} \label{sec:wsamodel}

As a third approach to deriving the background solar wind speed for input to ELEvoHI, we test the usage of the Wang-Sheeley-Arge and Heliospheric Upwind eXtrapolation models (WSA-HUX).
More specifically, we use magnetic maps of the photospheric field from the Global Oscillation Network Group (GONG) of the National Solar Observatory (NSO) as input to magnetic models of the solar corona. Using the Potential Field Source Surface model~\citep[PFSS;][]{altnew69, scha69} and the Schatten Current Sheet model~\citep[SCS;][]{scha71}, we compute the global coronal magnetic field topology. While the PFSS model attempts to find the potential magnetic field solution in the corona with an outer boundary condition stating that the field is radial at the source surface at 2.5 R$_\odot$, the SCS model accounts for the latitudinal invariance of the radial magnetic field in the region between 2.5 and 5 R$_\odot$ as observed in Ulysses field measurements~\citep[][]{wan95}. From the global magnetic field topology, we calculate the solar wind conditions near the Sun using the Wang-Sheeley-Arge model~\citep[WSA;][]{arg03}. To map the solar wind solutions from near Sun to Earth, we use the Heliospheric Upwind eXtrapolation model~\citep[HUX;][]{ril11b}, which simplifies the fluid momentum equation as much as possible. The HUX model solutions match the dynamic evolution predicted by global heliospheric MHD codes fairly well while having low processing power requirements. More details on the numerical framework can be found in \citet{rei19}.

\begin{figure}[h]
    \centering
    \includegraphics[width=15cm,keepaspectratio]{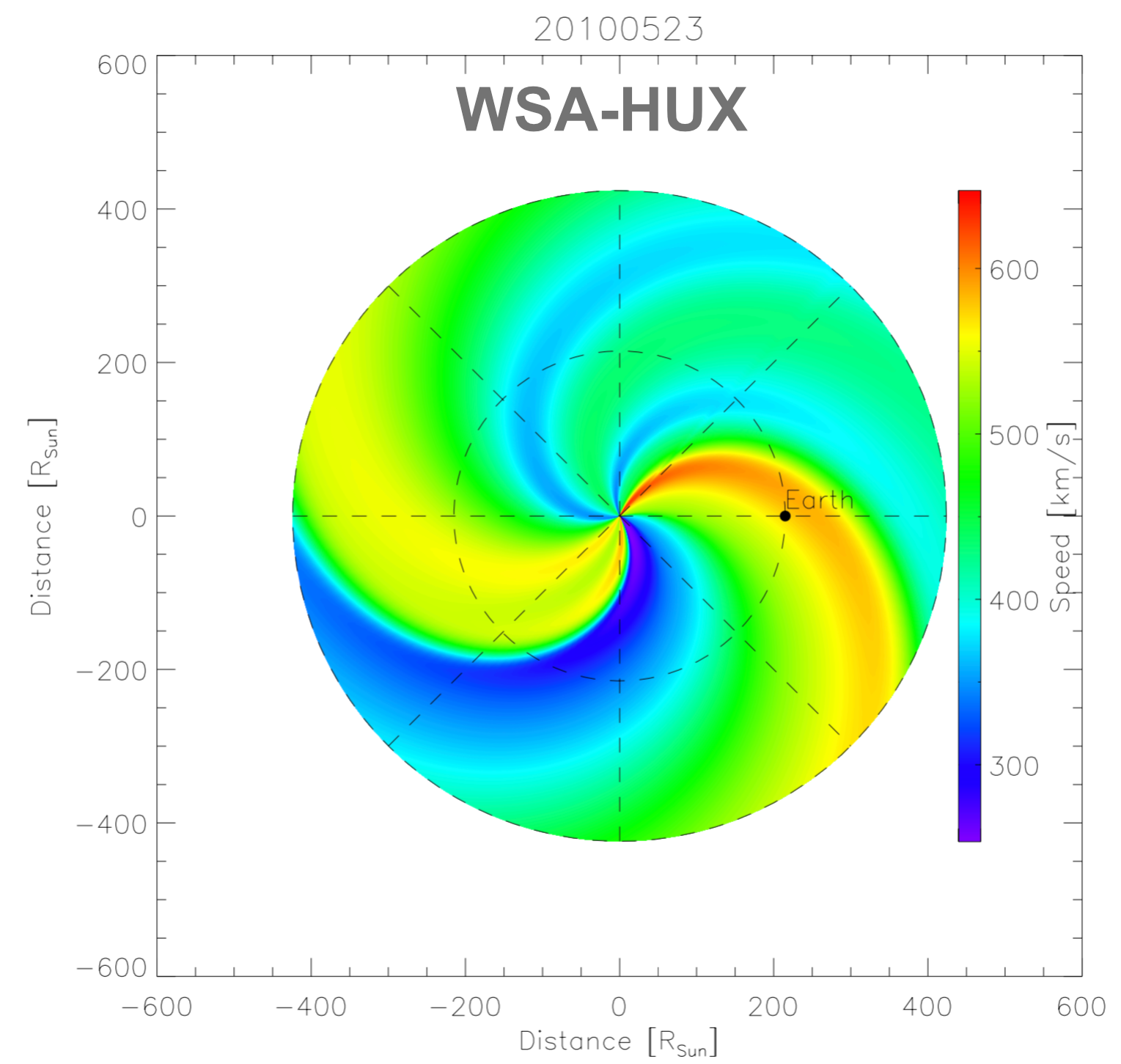}
    \caption{Example of the ambient solar wind speed for one event under study, observed by STEREO-A and STEREO-B (event n$^\circ$ 9 and n$^\circ$ 10 in Table \ref{tab:table0}). The region of interest is extracted and averaged and serves as input to the ELEvoHI ensemble model.}
    \label{fig:wsahux}
\end{figure}

Figure \ref{fig:wsahux} presents the modeled ambient solar wind for one event under study. For this method, we consider only that radial range of the heliosphere where the DBM fit is performed, i.e.\ between the two red vertical lines indicated in Figure \ref{fig:elcon} (approximately 30--100 R$_\odot$). In longitude, we use a range $\phi \pm \lambda$ to define the area in which the solar wind is acting on a certain CME ensemble member. The median value of the solar wind speed within this sector is calculated and a range of $\pm 100$ km~s$^{-1}$ is assumed. Over this range of ambient solar wind speeds, in steps of 25 km~s$^{-1}$, 9 DBMfits are performed.

To test the ELEvoHI model set-up, we use the three previously discussed methods to provide the source for the ambient solar wind speed, i.e.\
\begin{enumerate}
\item in situ data from 1 AU,
\item speed range derived from statistics,
\item modeled by WSA-HUX model.
\end{enumerate}

\subsection{Definition of the `optimal' DBM fit}

In the current version of ELEvoHI, the optimal DBM fit (out of several fits performed based on a range of input ambient solar wind speeds, as discussed in the previous section) is defined as the fit with the smallest mean residual to the time-distance profile along the whole extent of the fitted curve. The ambient solar wind speed associated with the best DBM fit is then used for further modeling. Usually, the DBM fit is performed over a radial distance of around 30 to 100 R$_\odot$. Sometimes we find that the DBM best fit does not actually agree very well with the last fitted data points, which can have a significant influence on the prediction. Therefore, it is tested if using only the mean residual of the last three fitted points leads to a better prediction than considering the residuals of the whole fit. Note that, in both cases, the total number of data points that are fitted stays the same, i.e.\ the track is fitted between the two end points that are manually chosen (vertical red lines in Figure \ref{fig:timespeedbox}). Only the evaluation of the residual differs in these two approaches.

For testing the ELEvoHI model set-up, we use the two previously discussed methods for evaluating the DBM fit and choosing the most suitable background solar wind speed, i.e.\
\begin{enumerate}
\item the smallest mean residual along the whole extent of the fit,
\item the smallest mean residual of the last three fitted points.
\end{enumerate}

\subsection{Benchmark model} \label{sec:fpf}

In order to compare the results of the different ELEvoHI ensemble runs to a well-established but simple prediction method that relies on HI data only, we use Fixed-Phi fitting \citep[FPF;][]{kahweb07,rou08}. The FPF method is the simplest of all such techniques based on HI data. It reduces the CME front to a point-like feature and assumes a radial propagation direction at a constant propagation speed. The best-fit equation of motion to the time-elongation profile extracted from HI data provides an estimate of the arrival time and speed at the target of interest. We apply FPF to the same time-elongation profiles as ELEvoHI and limit the track length to the start and end points between which the DBM fit is performed (red lines in Figure \ref{fig:elcon}), i.e.\ the same number of data points is used.
Although the method is simple, its predictions are not significantly worse than the predictions from more sophisticated methods \citep[][]{moe14}.
Using results from a benchmark model as a comparison provides the possibility to prove whether ELEvoHI is able to increase the prediction accuracy compared to the simple FPF method.

\section{Results} \label{sec:results}

\begin{figure}[h]
    \centering
    \includegraphics[width=15cm,keepaspectratio]{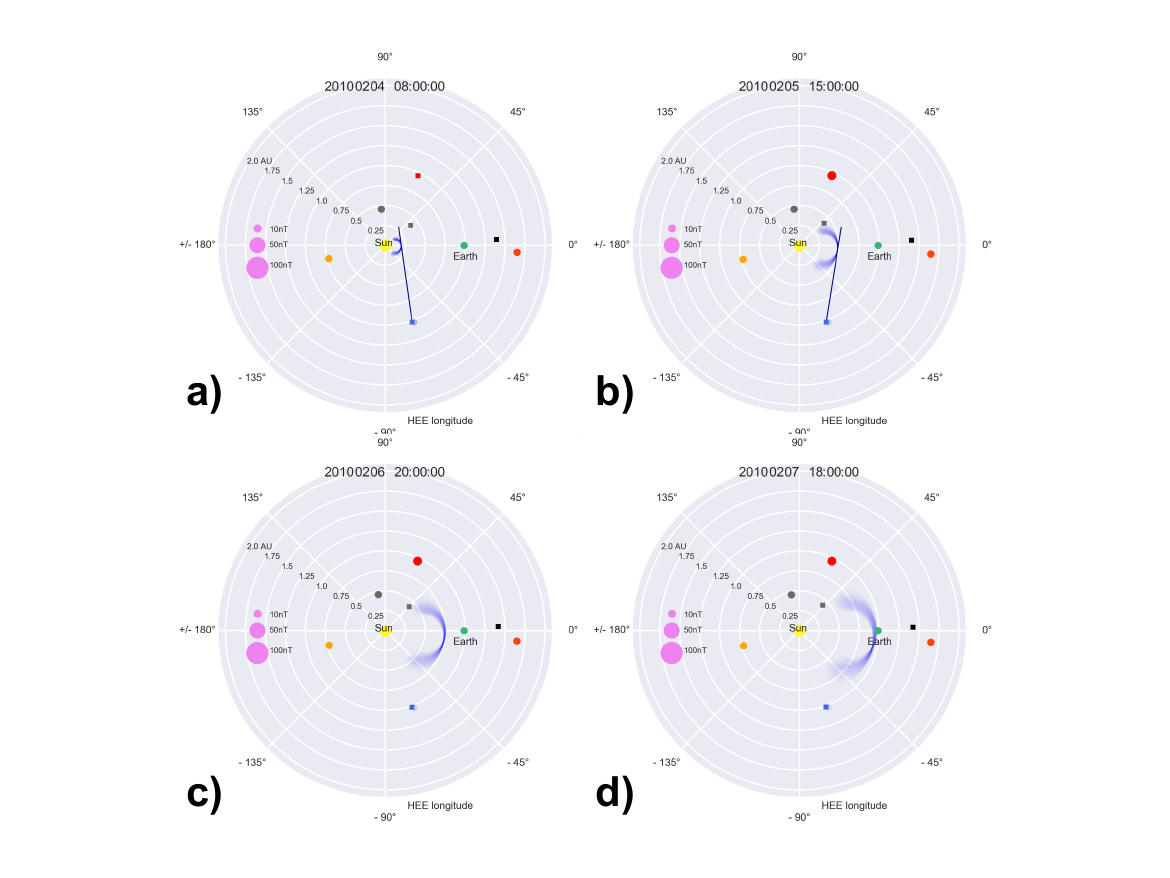}
    \caption{Four different time steps during ELEvoHI CME modeling for one example event (event n$^\circ$ 5 in Table \ref{tab:table1}). Panel a) shows the CME at the start of the DBM fit; the blue tangent corresponds to the measured HI-elongation observed from STEREO-B. Panel b) shows the end of the DBM fit and the last HI-elongation measurement used for the prediction. Panels c) and d) show additional times during CME evolution, while the latter corresponds to the time at which the CME was detected in situ at Earth. All movies and modeling results are available at \url{https://doi.org/10.6084/m9.figshare.12333173.v1}.}
    \label{fig:movie}
\end{figure}

We perform 18 ensemble runs for each CME in our list of 15 events by combining three different approaches related to the ambient solar wind speed, three different ways of gaining the CME frontal shape/direction and two different methods of defining the best DBM fit. Every ensemble run consists of 220 ensemble members (resulting from varying $\lambda$, $f$ and $\phi$ input parameters within certain ranges), i.e.\ for each event, we perform 3960 predictions with 59400 predictions in total. We calculate the median, the mean and the standard deviation of the distribution of predictions of the arrival time at Earth for each of the 18 ensembles and for each of the 15 CMEs. Figure \ref{fig:movie} shows four different time steps of the ELEvoHI simulation result for one example event (n$^\circ$ 5 in Table \ref{tab:table1}). Panels a) and b) correspond to the start and end times of the DBM fit; the blue tangent represents the corresponding HI elongation measurement. Panels c) and d) show later time steps of the prediction, for which no HI data were used (hence no blue tangent). Panel d) presents the time of the in situ arrival detection.

Overall, ELEvoHI achieves a mean absolute error (MAE) of $8.2 \pm 5.5$ h, a root mean square error (RMSE) of $11.1$ h, and a mean error (ME) of $+0.8$ h, the latter indicating that our model neither over- nor underestimates the transit time. Here, it is important to emphasize that these values arose from analyzing a small set of 15 isolated (non-interacting) events that were not predicted in real-time. Therefore, one has to be careful when comparing these results with the results of studies dealing with real-time predictions. In the following paragraphs, we evaluate the performance of the different model set-ups.

Table \ref{tab:table1} lists the MAE, the ME, the RMSE and the mean standard deviation (MSTD) of the difference between predicted and observed arrival time, $t$, and speed, $v$, for each of the 18 different model set-ups. Negative values correspond to an underestimated transit time, hence the event was predicted to arrive earlier than it actually did. In the case of the arrival speed prediction, negative values correspond to an underestimated arrival speed. The results are ordered from smallest to largest MAE in arrival time, revealing that the six set-ups using the WSA-HUX output as input for the ambient solar wind estimate lead to the most accurate predictions. The benchmark FPF technique leads to an MAE of 7.8 h with an MSTD of 10.5 h and an ME of 2.6 h, which means that FPF has a tendency to overestimate the transit time. Considering the underlying geometry assumed by FPF, this is not surprising. FPF reduces the CME front to a single point and assumes this point is tracked throughout the CME's propagation. Being conscious of the fact that CMEs can be extremely large-scale structures, it is clear that this is an oversimplification. Additionally, our CME sample almost exclusively consists of slow CMEs for which the assumption of constant propagation speed is usually close to reality. The faster the CME and hence the larger its likely deceleration, the larger the error due to a constant speed assumption \citep[][]{lug11}. However, we cannot dismiss the result that FPF performs as well as ELEvoHI (when averaging over all model set-ups) for the chosen set of CMEs. Again, it can be shown that a more sophisticated method is no guarant of a better prediction as already demonstrated by \citet{vrs14}, who compared the performance of the DBM and the WSA-Enlil+Cone model based on a list of 50 CMEs. The authors found that the two methods predicted the CME arrival time with an MAE of 14.8 and 14.1 h, respectively (for real-time predictions). Fortunately, this does not mean that we have already reached the best possible prediction accuracy; improving a method can still reap rewards. As Table \ref{tab:table1} shows, ELEvoHI based on phi from FPF can outperform the benchmark FPF when part of a more sophisticated model set-up, e.g. when coupled with WSA-HUX as the source of the solar wind.

\begin{figure}[h]
    \centering
    \includegraphics[width=10cm,keepaspectratio]{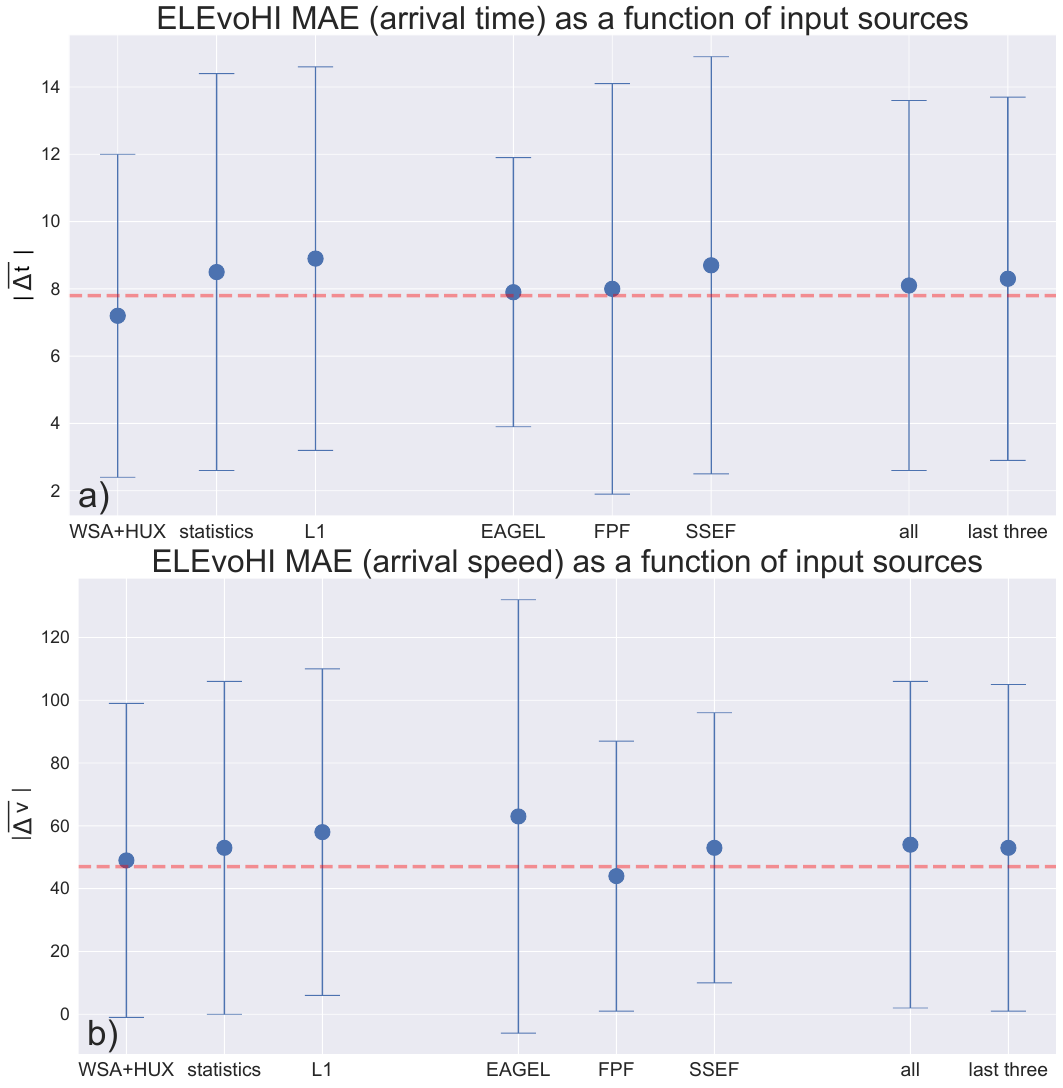}
    \caption{ELEvoHI MAE of a) arrival time and b) speed prediction corresponding to each source of input parameter. The left set of bars correspond to the three different sources of ambient solar wind input, the middle set of bars correspond to the three different sources of propagation direction (and CME frontal shape in case of EAGEL), and the right set of bars show the results that correspond to the two different ways of defining the best DBM fit. The error bars mark the standard deviation of the predictions. The horizontal dashed lines represent the MAE of the benchmark model, FPF.}
    \label{fig:source_comp}
\end{figure}

Figure \ref{fig:source_comp} shows the performance of the different model set-ups, grouped by the input type. The left/middle/right bars show the MAE and the MSTD of predictions based on different kinds of solar wind input/frontal shape/direction/best DBM fit. For all runs that use WSA-HUX, we find an MAE of $7.2 \pm 4.8$ h, using statistical background wind information results in an MAE of $8.5 \pm 5.9$ h and using the in situ solar wind speed from L1 we find an MAE of $9.0 \pm 5.7$ h. We find that using WSA-HUX as the background solar wind source results in 6 events being predicted better (i.e.\ with the smallest MAE) than using other solar wind sources (statistical background wind: 5, L1: 4). For one event in the list, no DBM fit converged when using WSA-HUX as input and therefore, no prediction was possible. The reason for that is that the solar wind speed range derived from WSA-HUX was not compatible with the HI kinematics. This can be the case when the HI data imply CME deceleration while the provided ambient solar wind speed is larger than the CME speed. In this situation, no physical solution relating to drag can be found as the CME cannot be decelerated below the speed of the ambient solar wind flow. This demonstrates the additional value of the whole approach because of the way that it avoids inappropriate values for the background solar wind speed. This is a distinct advantage over methods that only rely on coronagraph data and provide no possibility to validate the background solar wind speed used in the model.

\begin{sidewaystable}
\caption{Accuracy of each model set-up, sorted by the mean absolute error (MAE). MAE, the mean error (ME), the root mean square error (RMSE) and the mean standard deviation of the arrival time (t) and speed (v) prediction are given. The last column lists the corresponding model set-up indicating the inputs for direction (and shape in case of EAGEL), solar wind and the way of defining the best DBM fit.}
\label{tab:table1}
\centering
\begin{tabular}{rrrrrrrrr}
\hline
MAE(t) [h] & ME(t) [h] & RMSE(t) [h] & MSTD(t) [h]& MAE(v) [km~s$^{-1}$] & ME(v) [km~s$^{-1}$] & RMSE(v) [km~s$^{-1}$] & MSTD(v) [km~s$^{-1}$] & model set-up  \\
\hline
6.2 & 0.1 & 7.9 & 5.2 & 39.0 & 17.8 & 46.7 & 39.3 & FPF\_WSA-HUX\_all\\
6.4 & -0.1 & 8.0 & 5.1 & 36.8 & 18.6 & 45.1 & 38.6 & FPF\_WSA-HUX\_last\\
7.3 & 1.5 & 9.2 & 5.6 & 49.3 & 23.6 & 56.6 & 40.5 & SSEF\_WSA-HUX\_all\\
7.3 & -2.5 & 10.0 & 3.5 & 62.1 & 34.0 & 74.5 & 70.2 &  EAGEL\_WSA-HUX\_all\\
7.4 & 1.3 & 9.1 & 5.6 & 47.7 & 22.2 & 55.3 & 40.7 & SSEF\_WSA-HUX\_last\\
7.5 & -2.8 & 10.1 & 3.5 & 60.8 & 35.3 & 73.4 & 70.4 & EAGEL\_WSA-HUX\_last\\
7.7 & -1.8 & 10.3 & 4.5 & 65.0 & 33.5 & 75.3 & 71.0 & EAGEL\_stats\_all\\
7.8 & -2.1 & 10.3 & 4.6 & 62.7 & 35.8 & 72.9 & 71.7 & EAGEL\_stats\_last\\
8.0 & -2.1 & 10.1 & 3.9 & 61.9 & 34.9 & 75.0 & 64.6 & EAGEL\_L1\_all\\
8.0 & -2.1 & 10.2 & 3.9 & 62.4 & 34.4 & 75.3 & 64.7 & EAGEL\_L1\_last\\
8.6 & 2.7 & 12.2 & 6.6 & 45.5 & 11.2 & 55.8 & 43.9 & FPF\_stats\_all\\
8.8 & 2.6 & 12.2 & 6.6 & 43.4 & 11.4 & 54.5 & 43.5 & FPF\_stats\_last\\
8.9 & 3.3 & 11.8 & 6.5 & 51.7 & 17.2 & 62.9 & 42.8 & SSEF\_stats\_all\\
9.1 & 3.2 & 11.8 & 6.5 & 49.6 & 15.4 & 61.1 & 44.5 & SSEF\_stats\_last\\
9.1 & 2.4 & 13.1 & 6.7 & 51.0 & 15.7 & 63.9 & 45.8 & FPF\_L1\_all\\
9.1 & 2.4 & 13.1 & 6.6 & 50.3 & 15.5 & 63.6 & 45.4 & FPF\_L1\_last\\
9.8 & 2.7 & 13.0 & 6.5 & 61.0 & 25.6 & 80.4 & 45.7 & SSEF\_L1\_all\\
9.9 & 2.6 & 13.0 & 6.5 & 61.2 & 25.4 & 80.4 & 46.2 & SSEF\_L1\_last\\
\hline
\multicolumn{9}{l}{The benchmark model FPF results for the arrival time prediction in an MAE of $7.8$ h, an ME of $2.6$ h, an MSTD of $10.5$ h and a RMSE of $10.1$ h.}\\
\multicolumn{9}{l}{For the arrival speed prediction FPF results in an MAE of $X$ h, an ME of $X$ h, an MSTD of $X$ h and a RMSE of $60$ h.}
\end{tabular}
\end{sidewaystable}

Comparing the predictions based on different sources of CME frontal shape/direction input, we find that the input from the EAGEL tool leads to an MAE of $7.9\pm4.0$ h and predictions based on FPF and SSEF result in an MAE of $8.0\pm6.1$ h and $8.7\pm6.2$, respectively. Interestingly, predictions based on EAGEL result in a smaller MSTD (related to the arrival time) than those based on FPF or SSEF. This is the result of a smaller angular width derived by EAGEL for some events in the list than the value assumed for the predictions based on FPF and SSEF propagation direction. In terms of the number of best predictions, using EAGEL results in the most accurate arrival time predictions for 6 of the CMEs (SSEF: 5, FPF: 4).
The last comparison is made between the results of the two methods for defining the optimal DBM fit. Here, we find no significant difference between the results of the method that takes into account the residuals of the whole fit ($8.1\pm5.5$ h) and the method that uses residuals of the last three fitted points only ($8.2\pm5.4$ h). Nevertheless, using the whole fit for evaluation leads to the best prediction for 9 out of the 15 CMEs and using only the last three residuals leads to the best prediction for the other 6 CMEs. Contrary to our conclusion made above, this means that using the residuals over the whole fit has a clear advantage over using only the last three residuals to judge the fit.

\begin{figure}[h!]
    \centering
    \includegraphics[width=10cm,keepaspectratio]{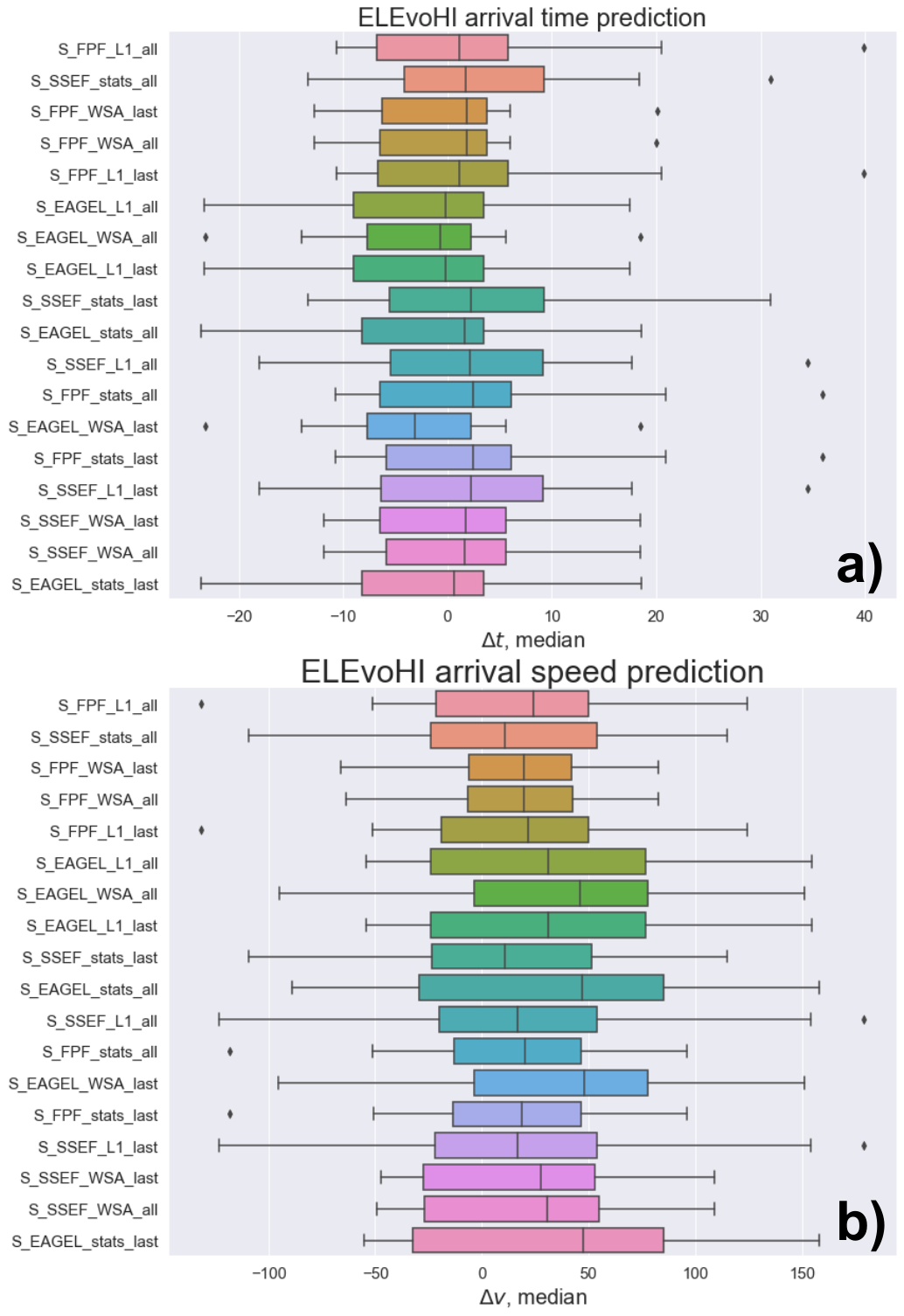}
    \caption{Overview of the prediction accuracy for every model set-up tested. Figure a) presents the prediction accuracy for the arrival time, figure b) for the arrival speed. The vertical lines within the boxes correspond to the median values, the boxes are delimited by the first and the third quartile, and the whiskers correspond to 1.5 times the interquartile range; the diamonds represent outliers.}
    \label{fig:timespeedbox}
\end{figure}

Evaluating ELEvoHI's ability to predict CME arrival speed based on each input parameter (Figure \ref{fig:source_comp}~b), we find that using WSA-HUX as background solar wind input, results in an MAE of $49\pm50$ km~s$^{-1}$ (L1: $58\pm52$ km~s$^{-1}$, statistical background wind: $53\pm53$ km~s$^{-1}$). Using input from EAGEL yields an MAE of $63\pm69$ km~s$^{-1}$ (FPF: $44\pm43$ km~s$^{-1}$, SSEF: $53\pm43$ km~s$^{-1}$). Judging the best DBM fit by the residuals of the whole fit gives $68\pm52$ km~s$^{-1}$ (last three residuals: $66\pm52$ km~s$^{-1}$). In case of the CME arrival speed prediction, the set-up used appears to make little difference.

Figure \ref{fig:timespeedbox} a) shows an overview of the performance of all of the different model set-ups as box and whiskers plots, based on the difference between predicted and actual arrival time for all events and all runs ($\sim3000$ runs per box). For almost every set-up the median is quite close to zero. This shows us that ELEvoHI has no bias towards providing predictions that are either too early or too late. As noted above, the overall RMSE($t$) is 11 h and the MAE($t$) is $8.2 \pm 5.5$ h, reflecting the actual prediction accuracy. The overall RMSE($v$) is 66 km~s$^{-1}$ and the MAE($v$) is 53 km~s$^{-1}$. Figure \ref{fig:timespeedbox} b) shows the analogous plot for arrival speed. Overall, ELEvoHI provides an MAE in the arrival speed prediction of $53\pm51$ km~s$^{-1}$, an RMSE of $66$ km~s$^{-1}$, and ME of $23$ km~s$^{-1}$. The latter means that ELEvoHI is not biased towards producing arrival speed predictions that are either too fast or too slow.

% ELEvoHI_MAE
% 8.203129895764777
% ELEvoHI_ME
% 0.7609645992693469
% ELEvoHI_std
% 11.03235439457074
% ELEvoHI_meanstd
% 5.505092330318882
% ELEvoHI_RMSE
% 11.058567294579593
% *************
% ELEvoHI_MAE_v
% 53.1496062992126
% ELEvoHI_ME_v
% 23.287401574803148
% ELEvoHI_std_v
% 61.57747567553432
% ELEvoHI_meanstd_v
% 51.01766529726526
% ELEvoHI_RMSE_v
% 65.83379514107602

\begin{figure}[h!]
    \centering
    \includegraphics[width=12cm,keepaspectratio]{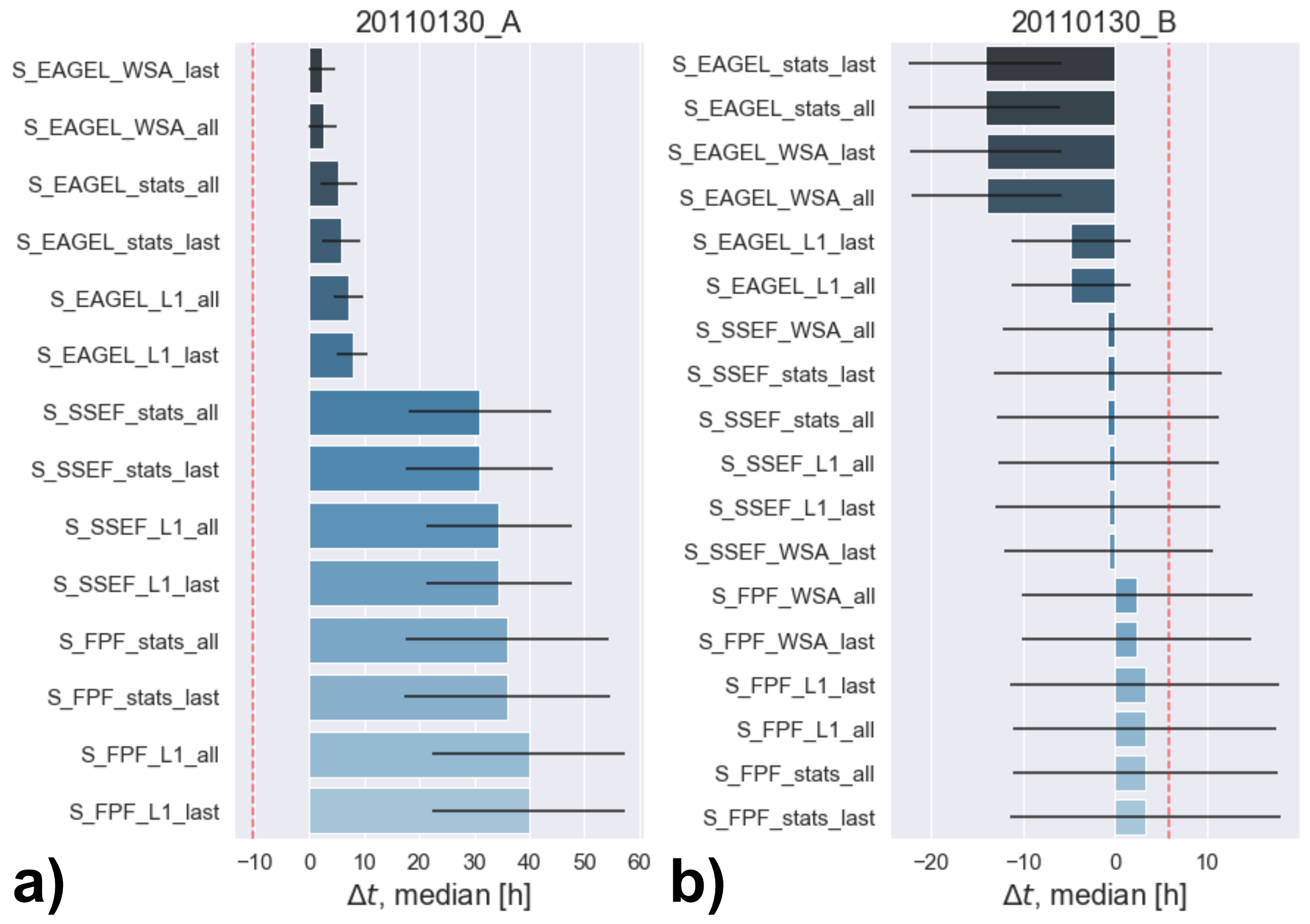}
    \caption{Comparison of ELEvoHI predictions for one example event (n$^\circ$ 14 and 15 in Table \ref{tab:table0}) performed separately for the two different vantage points, i.e.\ from STEREO-A and B, respectively. The dashed red line shows the result of the benchmark model.}
    \label{fig:stereo_comp}
\end{figure}

Some of the events under study (n$^\circ$4 \& 5, 9 \& 10, 14 \& 15 in Table \ref{tab:table0}) are observed and predicted from two separate vantage points, namely from STEREO-A and STEREO-B. Provided that the assumption of an elliptical self-similarly expanding structure is true, we might suppose that the predictions of an event observed from different sides agree well. Figure \ref{fig:stereo_comp} refutes this assumption by presenting the arrival time predictions for a CME launched on 2011-01-30. In panel a) the predictions based on STEREO-A/HI data (n$^\circ$ 14 in Table \ref{tab:table0}) are shown and panel b) presents the predictions based on STEREO-B/HI data (n$^\circ$ 15 in Table \ref{tab:table0}). Interestingly, the results are highly dependent on the model set-up used. For the view from STEREO-A, EAGEL+WSA-HUX input seems to be the best choice compared to the predictions based on the two HI fitting methods that lead to an error between 30 and 40 h. The combination of SSEF/FPF and WSA-HUX was not possible for this event from the vantage point of STEREO-A because the ambient solar wind speed range provided by WSA-HUX did not agree with the HI kinematics. Contrariwise, from the vantage point of STEREO-B, the EAGEL+WSA-HUX set-up leads to an error of more than 10 hours, while the predictions based on input directions derived from SSEF almost exactly match the in situ arrival time. A more detailed analysis on CMEs observed stereoscopically and modeled by ELEvoHI will be presented in a study by Hinterreiter et al.\ (in preparation for Space Weather).

This comparison shows that the current assumptions within ELEvoHI, i.e.\ constant ambient solar wind speed and elliptical CME frontal shape, are not correct for every event. When the CME is observed and predicted from the two different vantage points, the results can differ significantly; with the correct assumptions in place for a specific CME, this should not be the case. Therefore, including a deformable shape within ELEvoHI to simulate CME interaction with structures in the ambient solar wind might lead to an improvement of the predictions.
Indeed, observations from more than one vantage point could be used to help constrain the shape and kinematics of the CME leading to such an improvement in the arrival prediction accuracies. This finding supports the benefit of having HI observations from two separate vantage points, e.g.\ L1 and L5.

\section{Applicability for real-time predictions} \label{sec:realtime}

By far the fastest and according to our findings in this study, a relatively satisfying way to set-up ELEvoHI, is using a combination of FPF and the statistical ambient solar wind approach. FPF uses the same data as needed by ELEvoHI, i.e.\ the HI time-elongation track. The FPF fitting method yields the propagation direction needed by ELEvoHI, while the half-width within the ecliptic plane can be assumed to be between 30 and 50$^\circ$ (it can be assumed, indeed, to be any other value). The statistical solar wind approach is directly implemented within the ELEvoHI model. As shown above, this set-up leads to an MAE in arrival time of 8.6 h and an ME of 2.7 h. However, if an ambient solar wind solution is available in real-time (e.g. the WSA-HUX or similar), ELEvoHI can achieve an MAE of 6.2 h with an ME of 0.1 h---still without the necessity for additional coronagraph data or the need for manual fitting to these images. Of course, we always need to keep in mind that these values are derived from a pre-defined set of very well-observed, and isolated, events and from HI science quality data that is currently not available in real-time. However, HI beacon data is available in near real-time and can serve as input to ELEvoHI since STEREO-A/HI is already close to L5 and is observing the space between the Sun and Earth---hopefully until 2027, when it will be around L4.
An additional possibility for having HI real-time data available in the future might be provided by the Polarimeter to Unify the Corona and Heliosphere (PUNCH) mission. PUNCH will be launched in 2023 and will operate in low Earth orbit.

For real-time predictions, it is of the utmost importance to be able to include an estimate of the arrival probability with a CME prediction. Currently, ELEvoHI simply calculates this as the ratio of the number of ensemble members that are predicted to hit the target to the total ensemble size. This is going to be updated in the near future, to give predicted flank hits a lower weighting. In addition, we have noticed that for flank hits, the arrival time error tends to be larger than expected and the transit time is overestimated. This could be due to the elliptical shape of the front resulting in highly curved flanks. In the future, we will examine if we can find a suitable approach to deal with these strongly bent flanks to avoid such extreme delays when predicting a flank encounter.

\section{Summary and Conclusions} \label{sec:conclusions}

In this work we studied 18 different combinations of inputs to run the HI-based ensemble CME arrival prediction model, ELEvoHI, in order to ascertain the set-up leading to the most accurate arrival time and speed predictions. As input for the ambient solar wind that influences the drag-based propagation of the modeled CME we used 1) the WSA-HUX background solar wind model, 2) an approach of simply providing a range of possible solar wind speeds (225--625 km~s$^{-1}$) derived from 14 years of observations at L1, and 3) the solar wind speed measured in situ at L1 during the evolution of the CME. We found that having a more accurate ambient solar wind as input leads to significantly better arrival time prediction. Using input from WSA-HUX improves the MAE by an hour, compared to simply providing a range for solar wind speeds, and leads to almost two hours improvement on MAE over the usage of L1 solar wind speed.

To analyze the influence of the CME frontal shape/propagation direction on ELEvoHI predictions, we compared three different sources of $\lambda$ and $\phi$: 1) Coronagraph images were used to perform a GCS-fit to derive the 3D shape of the CME. The intersection of this 3D front with the ecliptic plane provides a 2D structure from which the measured angular half-width and direction were input to ELEvoHI. 2) The FPF and 3) the SSEF HI fitting methods, which only provide the direction of motion. In these cases, we had to assume a half-width (we chose a range between 30 and 50$^\circ$). In all cases we had to assume $f$ to vary between $0.7$ and 1, while a value of 1 corresonds to a circular frontal shape. Surprisingly, approach 1 did not lead to a significantly more accurate prediction than using FPF or SSEF and simply assuming the half-width to lie within a certain range. One possible reason for this might be different direction of motion within the coronagraph field of view compared to that within the HI field of view. Another reason might be an ongoing rotation of the CME, leading to a change in its angular half-width within the ecliptic. However, this is a surprising result as one would expect a more data-oriented input (regarding the CME shape) to lead to better predictions.

The third aspect of the model set-up that we tested is the model-intrinsic procedure to determine the optimal DBM fit to the HI kinematics. This process defines the ambient solar wind speed, which is further used as a basis for the arrival prediction. Several DBM fits were performed over the provided range of solar wind speeds, the best DBM fit then determined the optimal ambient solar wind speed and drag parameter. We compared two ways of defining the best DBM fit, namely the fit with the minimum value of the mean residual of 1) the whole fit or 2) the last three points of the fit. We found that both procedures lead to similar results, but with using the residuals of the whole fit leading to slightly better predictions.

Based on this study, we are now able to operate ELEvoHI to gain the best possible arrival predictions. Our results emphasize the importance of an accurate ambient solar wind model, as the solar wind heavily influences the drag-based evolution of the CME. In the future, an interesting advancement might be to include a range of values for the solar wind to contribute to the ensemble instead of deriving only a single value per shape/direction set-up. Another logical next step would be to release ELEvoHI from its rigid elliptical shape and to allow deformation due to the influence of the ambient solar wind. In any case, with ELEvoHI, we are prepared for real-time CME arrival predictions, once a new HI observer is delivering high quality data in real-time.

\section{Data Sources}

\noindent \textbf{Data} \\

\noindent STEREO/HI: \url{https://www.ukssdc.ac.uk/solar/stereo/data.html} \\
\noindent STEREO/COR2 and SoHO/LASCO: \url{https://sdac.virtualsolar.org/cgi/search} \\
\noindent NSO/GONG: \url{https://gong.nso.edu/data/magmap/}

\noindent \textbf{Model} \\

\noindent ELEvoHI is available on github under \url{https://github.com/tamerstorfer/ELEvoHI/releases/tag/v1.0.0.0}.

\noindent \textbf{Results} \\

\noindent The visualization of each prediction result, i.e.\ movies and figures, can be downloaded from \url{https://doi.org/10.6084/m9.figshare.12333173.v1}.

%%%%%%%%%%%%%%%%%%%%%%%%%%%%%%%%%%%%%%%%%%%%%%%%%%%%%%%%%%%%%%%%
%
%  ACKNOWLEDGMENTS
%
% The acknowledgments must list:
%
% >>>>  A statement that indicates to the reader where the data
%   supporting the conclusions can be obtained (for example, in the
%   references, tables, supporting information, and other databases).
%
%   All funding sources related to this work from all authors
%
%   Any real or perceived financial conflicts of interests for any
%   author
%
%   Other affiliations for any author that may be perceived as
%   having a conflict of interest with respect to the results of this
%   paper.
%
%
% It is also the appropriate place to thank colleagues and other contributors.
% AGU does not normally allow dedications.

\acknowledgments
%Enter acknowledgments, including your data availability statement, here.
T.A., J.H., M.B., M.R., C.M., A.J.W., R.L.B, and U.V.A. thank the Austrian Science Fund (FWF): P31265-N27, J4160-N27, P31659-N27, P31521-N27. M.D. acknowledges support by the Croatian Science Foundation under the project 7549 (MSOC).

%% ------------------------------------------------------------------------ %%
%% References and Citations

%%%%%%%%%%%%%%%%%%%%%%%%%%%%%%%%%%%%%%%%%%%%%%%
% BibTeX is preferred:
%
% \bibliography{<name of your .bib file>}
%\bibliography{tam_bib}

% don't specify bibliographystyle
% \begin{thebibliography}{}

\end{document}